# Multimodal Emotion Coupling via Speech-to-Facial and Bodily Gestures in Dyadic Interaction

Von Ralph Dane Marquez Herbuela
International Research Center
for Neurointelligence (WPI-IRCN)
The University of Tokyo
Tokyo Japan
herbuela.vonralphdane@mail.u-tokyo.ac.jp

Yukie Nagai
International Research Center
for Neurointelligence (WPI-IRCN)
The University of Tokyo
Tokyo Japan
nagai.yukie@mail.u-tokyo.ac.jp

## Abstract

Human emotional expression emerges through coordinated vocal, facial, and gestural signals. While speech–face alignment is well-established, the broader dynamics linking emotionally expressive speech to regional facial and hand motion—remains critical for gaining a deeper insight into how emotional and behavior cues are communicated in real interactions. Further modulating the coordination is the structure of conversational exchange like sequential turn-taking, which creates stable temporal windows for multimodal synchrony, and simultaneous speech, often indicative of high-arousal moments, disrupts this alignment and impacts emotional clarity. Understanding these dynamics enhances real-time emotion detection by improving the accuracy of timing and synchrony across modalities in both human interactions and AI systems. This study examines multimodal emotion coupling using region-specific motion capture from dyadic interactions in the IEMOCAP corpus. Speech features included low-level prosody, MFCCs, and model-derived arousal, valence, and categorical emotions (Happy, Sad, Angry, Neutral), aligned with 3D facial and hand marker displacements. Expressive activeness was quantified through framewise displacement magnitudes, and speech-to-gesture prediction mapped speech features to facial and hand movements. Non-overlapping speech consistently elicited greater activeness particularly in the lower face and mouth. Sadness showed increased expressivity during non-overlap, while anger suppressed gestures during overlaps. Predictive mapping revealed highest accuracy for prosody and MFCCs in articulatory regions while arousal and valence had lower and more context-sensitive correlations. Notably, hand–speech synchrony was enhanced under low arousal and overlapping speech, but not for valence. These highlight the role of conversational structure and emotional dimensions in shaping multimodal expression—emphasizing the need for region-aware models of emotion communication.

## 1 Introduction

Human communication is inherently multimodal, relying on the tightly coordinated interplay of speech, facial expressions, and gestures to convey not just linguistic content but emotional nuance and social intent. This phenomenon, referred to as multimodal synchronization, supports not only information transmission but also interpersonal rapport, empathy, and emotional resonance [1, 2, 3]. It is this fine-grained, moment-to-moment alignment across modalities that defines naturalistic emotional communication—and understanding its structure remains a central challenge in affective computing [4, 5].

One well-studied form of multimodal synchrony is the coupling of speech prosody with facial expressions. Temporal alignment between pitch contours, intensity shifts, and speech rhythm with facial gestures—such as eyebrow movements, eye blinks, and mouth shape—enhances emotional clarity, communicative precision, and improves the recognition of emotional states and facilitates deeper interpersonal connection [1, 2, 6, 7]. It plays a particularly important role in emotionally salient contexts, such as disagreement, persuasion, or empathy. Despite its centrality in human interaction, computational systems for emotion recognition have often treated these modalities independently, applying either coarse-grained averaging or ignoring their temporal dependencies altogether [8].

In parallel, hand and upper-body gestures—although long recognized as integral to co-speech communication—remain underrepresented in computational models of emotional synchrony. Gestures serve a range of communicative and affective functions: they can emphasize speech, mirror prosodic structure, reveal internal emotional states, and provide a channel for self-regulation. Co-speech gestures have been shown to temporally align with prosodic and semantic features of speech, enhancing the expressiveness, rhythm, and emotional clarity of interaction [9, 10, 11, 12]. When coordinated with vocal and facial signals, gestures contribute to a more embodied and behaviorally grounded portrayal of affect. This coordination has particular significance in neurodiverse and developmentally diverse populations, where gesture may amplify or compensate for less expressive verbal or facial signals [13, 14]. Despite these advances, most existing emotion recognition systems treat gesture as ancillary or model it separately from speech and facial cues, limiting our understanding of how multimodal emotion is expressed in real-time interaction. Recent research emphasizes the need for temporally precise and contextually integrated gesture modeling to improve affective computing frameworks [15, 16, 17].

Further modulating this coordination is the structure of conversational exchange—particularly the distinction between whether participants take turns or speak simultaneously shapes the timing and synchrony of expressive behavior. Non-overlapping speech—characterized by sequential turn-taking—provides structured temporal windows that facilitate more stable and analyzable patterns of multimodal synchrony [18, 19]. In contrast, overlapping speech, often linked to heightened arousal or engagement, may foster mutual responsiveness and affective mirroring [20, 21]. It reflects spontaneity and heightened involvement but can also introduce temporal noise that disrupts clean alignment across channels. Analyzing how different conversational structures influence temporal and cross-modal coherence is therefore essential for capturing the layered dynamics of natural emotional interaction.

This study presents a multimodal emotion analysis framework that builds on and significantly extends the audiovisual mapping paradigm introduced by Busso and Narayanan [22]. Specifically, we incorporate frame-level speech emotion modeling using deep neural networks, include both facial and hand gesture modalities, and analyze the influence of spontaneous conversational dynamics—including overlapping and non-overlapping speech—on multimodal expressivity. Our analysis is based on the IEMOCAP dataset, which provides rich dyadic recordings with synchronized audio, video, transcript, and motion capture data. By extracting prosodic and spectral speech features alongside categorical and dimensional emotion predictions, and aligning them with 3D facial and hand marker data segmented by region, we aim to characterize both the relationships between speech features and expressive behavior, and how facial and hand activeness varies systematically across emotional states and speech conditions.

Specifically, our contributions are as follows:

(1) Integration of categorical (SUPERB) and dimensional (audEERING) model-predicted emotional descriptors and prosodic and MFCC-based speech features to drive multimodal mapping.

(2) Segmentation of facial data into anatomically meaningful subregions (upper, middle, lower) and combine them with hand gesture features for region-specific analysis of expressive motion.

(3) Analysis of speech, facial, and hands motions under different interactional structures—non-overlapping vs. overlapping speech—to assess the influence of conversational dynamics on multimodal expressivity.

(4) Application of the Affine Minimum Mean Square Error (AMMSE) model for speech-to-motion mapping that yields interpretable, temporally aligned transformations between emotion-laden speech and facial/hand movement. This framework allows us to investigate how emotion-related speech features correspond with distinct patterns of facial and hand movement across segmented regions and conversational conditions, offering a detailed view of multimodal coordination in dyadic interaction, with particular emphasis on how these patterns vary across different emotional states.

## 2 Related Work

### 2.1 Speech-to-motion mapping

Mapping speech features to facial motion using AMMSE has enabled analysis of audiovisual synchronization during emotion expression. While some extensions explored correlations with prosody and gestures [23], other recent approaches have adopted neural architectures such as LSTMs and temporal convolutional networks (TCNs) to capture sequential dynamics in speech-to-motion modeling [24]. However, these methods have largely focused on generating motion for virtual agents rather than analyzing naturalistic interactions. Our study instead revisits AMMSE [22] to examine region-specific facial and hand dynamics during spontaneous dialogue. Few mapping studies have addressed subregional analysis or incorporated upper-body gestures, though co-speech gesture modeling has highlighted the expressive role of hand movement [9].

### 2.2 Speech structure in multimodal synchrony

Our prior work underscores the impact of conversational structure on emotional synchrony. Using the IEMOCAP dataset, we compared overlapping and non-overlapping speech segments and found that the latter condition yielded significantly higher alignment in arousal and valence trajectories across facial and vocal modalities. Temporal correlation and lag analyses further revealed shorter delays and stronger synchrony during non-overlapping speech, suggesting more immediate co-expression of affect across modalities. These findings provided quantitative evidence that conversational format modulates the strength and timing of multimodal emotional alignment, motivating the present study's focus on integrating facial, vocal, and gestural dynamics under different interaction conditions.

Despite advances in each of these areas, few studies have jointly modeled emotion-related speech features, facial and hand motion, and conversational structure at a high temporal resolution. Our work addresses this gap by integrating frame-level emotional analysis with region-specific MoCap dynamics under varied speech conditions.

## 3 Method

Our multimodal analysis pipeline (Figure 1) integrates synchronized audio, facial motion, and hand gestures from the IEMOCAP dataset to investigate how vocal emotion signals align with expressive bodily movements. Using audio extracted from the left-positioned speaker, we computed both low-level acoustic features (pitch, energy, MFCCs) and frame-level categorical and dimensional emotion estimates using pre-trained wav2vec2-based models. These were aligned with 3D facial and hand motion capture data segmented into upper, middle, and lower facial regions and hand trajectories. All modalities were temporally synchronized at 60.24 Hz and annotated for overlapping and non-overlapping speech conditions, enabling fine-grained analysis of multimodal coordination across speech, gesture, and emotion.

### 3.1 The IEMOCAP dataset

We conducted our analysis using the Interactive Emotional Dyadic Motion Capture (IEMOCAP) dataset [25], a multimodal corpus comprising five sessions of scripted and improvised dyadic interactions between professional actors. Each session features male-female pairs recorded with synchronized high-resolution video (30 Hz), stereo audio (48 kHz), utterance-level transcripts, and 3D motion capture (MoCap) of facial and upper-body gestures sampled at 120 Hz, allowing detailed spatial analysis of expressive behavior. IEMOCAP is widely used in affective computing due to its rich combination of spontaneous dialogue, varied emotional content, multimodal annotations, and coverage of overlapping and non-overlapping speech segments, making it particularly suitable for studying how interaction structure shapes multimodal synchrony.

### 3.2 Pre-processing

Each audiovisual recording in the IEMOCAP corpus contains stereo audio tracks corresponding to two speakers engaged in dyadic interaction. For this study, we isolated the speech signal of the left-positioned speaker, which served as the primary modality for predicting bodily gestures. Audio separation was conducted using FFmpeg's channelsplit filter, leveraging the known recording configuration: for Session 1, the Front Left (FL) channel was extracted, and for Sessions 2 through 5, the Front Right (FR) channel was used to maintain consistency in targeting the left speaker across sessions. To refine the dataset further, the initial four seconds of each audio file were trimmed using FFmpeg's stream copy functionality. This step removed preparatory or non-task behaviors typically present at the beginning of recordings, ensuring that analyses focused on active emotional speech segments.

### 3.3 Speech emotion extraction

In Busso's original IEMOCAP framework [25] emotional analysis primarily based on manually annotated by human raters at the utterance level. In contrast, our approach leverages model-predicted frame-level emotional estimates derived from pre-trained deep neural network models to captures continuous, high-resolution representation of fluctuations in emotional expression over time rather than assuming a single label per utterance. We employed the wav2vec2-base-superb-er model from the SUPERB (Speech Processing Universal PERformance Benchmark) benchmark suite [26]. The SUPERB framework was trained to classify utterances into four primary emotional categories: Neutral, Happy, Sad, and Angry. The wav2vec2-base-superb-er model uses the wav2vec 2.0 backbone architecture (Figure 2), a convolutional and transformer-based self- supervised model that learns high-level speech representations from raw audio. For the ER task, a classification head is added on top of the transformer outputs, which maps each segment to one of the emotion categories. Feature extraction was performed on a frame-by-frame basis, synchronized with our 100 ms segmentation windows to enable fine-grained emotional annotation throughout the interaction.

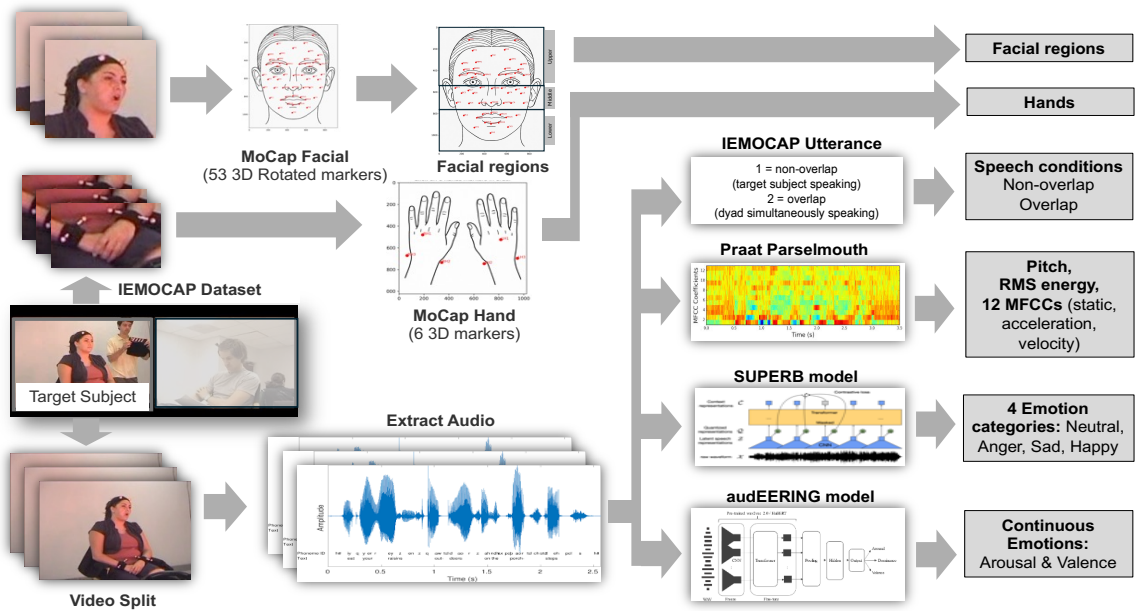

**Figure 1: Multimodal emotion coupling pipeline for gesture activeness and speech-to-facial and hand motion predictions.**

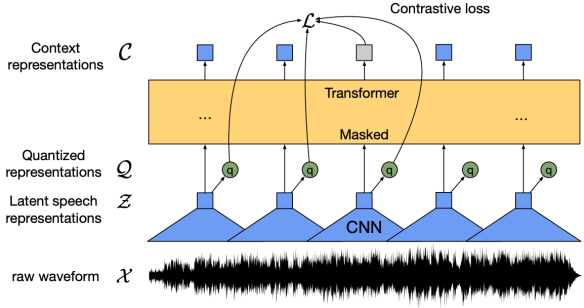

Figure 2: Overview of the wav2vec 2.0 speech emotion recognition architecture, adapted from Baevski et al. (2020) [27].

While categorical labels remain valuable for providing clear and interpretable distinctions between emotional states, they can be limited in capturing the nuanced and dynamic nature of emotion during spontaneous interaction. To address this, we additionally extracted continuous arousal and valence estimates using the wav2vec2-large-robust-12-ft-emotion-msp-dim model developed by audEERING [28]. Dimensional modeling represents emotion as a point in a continuous affective space, allowing for the expression of mixed, subtle, or transitional states that would otherwise be flattened into discrete categories. As shown in Figure 3, the audEERING model is a self-supervised transformer architecture fine-tuned on the MSP-Podcast corpus for dimensional emotion recognition [28]. Based on the pretrained wav2vec2 2.0 or HuBERT models which processes raw audio waveform through a convolutional feature encoder, and then projects the input into a high-dimensional latent space. These representations are passed through multiple transformer layers employing multi-head self-attention mechanisms to model temporal dependencies. The final output layer, custom regression head was trained to output frame-level arousal, valence, and dominance values. Audio signals were first resampled to 16kHz and segmented into overlapping 100ms windows. In this study, we focused only on arousal and valence, normalized to the range [-1, 1], as these two dimensions are more robustly supported across affective computing research and are better validated in the training data.

### 3.4 Prosodic and spectral speech features

To analyze the relationship between speech and body motion, we extracted low-level acoustic features that capture both prosodic and spectral characteristics of the speech signal. All features were extracted using the Praat software via the Parselmouth Python interface. Each audio waveform was first resampled to 16kHz and analyzed with a time step of 8.3 ms (120 Hz) and a 25 ms window size. The prosodic features included the fundamental frequency (F0), representing the pitch contour, and the root mean square (RMS) energy, reflecting the amplitude envelope of the speech waveform. F0 was extracted using Praat's `to_pitch` method with a pitch range of 50–500 Hz. RMS energy was extracted using `to_intensity`. Spectral information, Mel-Frequency Cepstral Coefficients (MFCCs), which model the short-time spectral envelope of the signal were extracted using the Praat's to_mfcc. A set of 13 MFCCs was initially computed, but the first coefficient (representing overall energy) was discarded to avoid redundancy with RMS energy, resulting in 12 retained coefficients. In addition to static features, we calculated the first and second temporal derivatives ($\Delta$ and $\Delta\Delta$) of all features—F0, RMS energy, and MFCCs to capture dynamic changes over time, as temporal dynamics have been shown to improve audio-visual mapping performance [22]. This resulted in a 42-dimensional feature vector per frame: 14 base features (F0, energy, 12 MFCCs) × 3 (static, $\Delta$, $\Delta\Delta$). To reduce the dimensionality of the spectral features while retaining most of the relevant information, we applied Principal Component Analysis (PCA) to the MFCC-derived features, following the approach proposed by Busso and Narayanan (2007) [22]. Twelve principal components were retained, capturing approximately 95% of the cumulative variance in the spectral feature space. The resulting 12 decorrelated spectral features were then concatenated with the six original prosodic features ((F0, energy) × 3 (static, $\Delta$, $\Delta\Delta$)), yielding a compact 18-dimensional speech feature vector per frame optimized for subsequent speech-to-gesture mapping analyses.

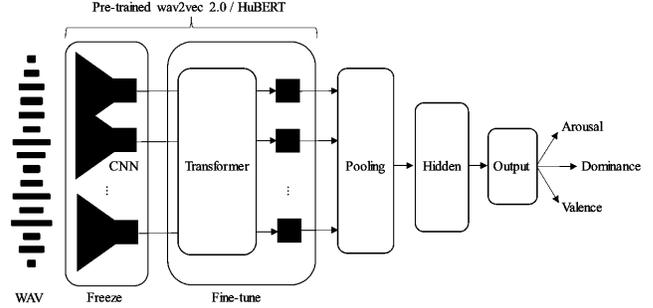

Figure 3: Speech emotion regression model architecture adapted from Wagner et al. (2022) [28].

### 3.5 Facial and hand motion feature extraction

In this study, we focused on two sets of markers from IEMOCAP: (1) rotated facial markers representing expressive dynamics after head motion compensation, and (2) hand markers representing gross upper-body gestures. The rotated MoCap files directly provide facial marker displacements that reflect expressive dynamics rather than rigid head motion. The rotated facial recordings contained 53 landmark points covering critical expressive areas including the eyebrows, eyes, nose, cheeks, and lips (Figure 4a). Upper face region includes markers above the eyes and on the forehead, associated primarily with emotional expressions conveyed by the brow and forehead: FH1, FH2, FH3 (forehead markers); LBM0–LBM3, RBM0–RBM3 (left and right brow movement); LBRO1–LBRO4, RBRO1–RBRO4 (left and right brow rotation), and LLID and RRID (left and right eye lids). Unlike Busso's approach, we retained the nose and nostril markers (MNOSE, TNOSE, RNSTRL, LNSTRL), because they now reflect dynamic midface expressions rather than rigid reference points, so these markers along with upper cheek areas, capturing mid-facial dynamics involved in emotional and articulatory expressions are part of the middle face region: LC2–LC8 (left cheek), RC2–RC8 (right cheek), MNOSE (mid-nose), TNOSE (tip-nose), LNSTRL (left nostril), and RNSTRL (right nostril) markers. Finally, lower face region includes markers below the nose, including the mouth contour, jawline, and lower cheeks, reflecting articulatory gestures and expressive mouth movements: MOU1–MOU8 (mouth region), CH1–CH3 (chin), LC1 (lower left cheek) and RC1 (lower right cheek).

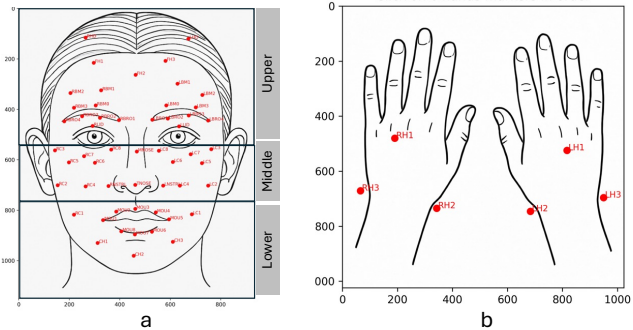

**Figure 4: Facial and hand markers placement for motion analysis.**

An important extension of our study compared to Busso and Narayanan [22] was the inclusion of hand motion analysis (Figure 4b). The IEMOCAP corpus provides three markers per hand (attached to the wrist [2 and 3] and hand [1]): RH1-3 (right hand) and LH1-3 (left hand). Hand motion features were extracted directly from the 3D trajectories of six hand markers (RH1–3 and LH1–3), which capture gross upper-limb motion over time. No additional kinematic features (e.g., velocity or acceleration) were derived for this analysis, as our aim was to preserve the raw spatial dynamics of gestural movement for multimodal alignment and correlation analysis. In Busso and Narayanan (2007) [22], facial motion features also included articulatory features, particularly eyebrow motion vectors and inter-lip distances based on Euclidean geometry between specific pairs of markers. Because the rotated MoCap files already isolate expressive movements by compensating for rigid head motion, we bypass additional rotation or alignment preprocessing. We utilize all available facial markers—including those along the eyebrows (e.g., LBM0-LBM3, RBM0-RBM3 for left and right brow movement), and lip regions (e.g., MOU1-MOU8 for mouth)—to compute framewise displacement magnitudes across predefined facial regions. Additionally, while head markers (left: LHD; right: RHD) are traditionally used for rigid head stabilization, we treat them as part of a separate region to quantify head activeness itself.

### 3.6 Temporal Alignment and Synchronization

After preprocessing each modality, all data streams were aligned to a 60.24 Hz frame structure. The original 30 Hz IEMOCAP video was upsampled to match higher-frequency data like MoCap (120 Hz) and prosodic features (extracted at 120 Hz, then downsampled). Emotion estimates from audEERING and SUPERB were upsampled from 10 Hz to 60.24 Hz. Prosodic features were downsampled to ~60.24 Hz by selecting every other frame, preserving dynamic patterns. MoCap was resampled from 120 Hz to 60.24 Hz using linear interpolation. Speaker-specific intervals from transcripts were discretized to 60.24 Hz and used to generate binary labels for speech and overlap. Finally, all data streams were merged into a single dataframe per session, synchronized by time, enabling analysis of emotional dynamics during speech.

### 3.7 Analysis

We first quantified expressive activeness by computing framewise 3D displacement magnitudes for each facial and hand marker. These magnitudes were averaged within predefined anatomical regions (e.g., upper face, mouth, hand) to derive region-level activeness per frame. From these, we generated dyad-level and group-level summaries segmented by speech condition (overlapping vs. non-overlapping) and emotion category (Happy, Sad, Angry, Neutral). To visualize expressive dynamics, we created anatomical heatmaps showing average regional activeness across conditions and emotions. Group-level differences were statistically tested using repeated-measures ANOVA, with emotion and speech condition as within-subject factors. This allowed us to examine how conversational structure and emotion jointly modulated facial and hand expressivity.

To investigate how acoustic signals predict expressive motion, we applied the AMMSE estimator. Separate AMMSE models were trained to map each feature set—prosody (pitch, energy), MFCCs, arousal, and valence—to motion across different facial and hand regions. For each region and feature, we computed Pearson's correlation between predicted and actual motion trajectories. These correlations were averaged across dyads and stratified by speech condition and affective bin (high/low arousal or valence). Finally, we visualized the AMMSE correlation results using anatomical heatmaps, enabling topographic comparison of acoustic–motion alignment across conditions. This multi-level approach revealed how expressive behavior is jointly shaped by interaction structure, emotion, and underlying acoustic dynamics.

## 4 Results and Discussion

### 4.1 Facial and hand activeness

Figure 5 presents regional activeness heatmaps across facial and hand markers for each emotion, separated by speech condition (non-overlapping vs. overlapping). Across all emotions, non-overlapping speech generally elicits stronger activeness, particularly in the upper and lower face and hands—highlighting the role of structured turn-taking in facilitating expressive coordination. In Angry interactions (Figure 5b), non-overlapping speech evokes broad activation across the brows, cheeks, and hands, whereas overlapping speech narrows this expressivity—maintaining upper-face intensity (notably brow and forehead), but with marked suppression in mid- and lower-face and nearly absent hand motion. This suggests that high-arousal conflict scenarios under overlap may constrain full-body expressivity, yielding a facially focused but gesture-suppressed profile.

In contrast, Happy (Figure 5a) and Sad (Figure 5c) states show fuller engagement during non-overlapping speech, with heightened activation in eyelids, mouth, and bilateral hand regions—supporting theories that positive and affect-rich dialogue engages holistic motor coordination. Overlapping speech reduces this distribution, especially around the cheeks and mouth, though some expressive facial motion remains. Neutral speech (Figure 5d), while often considered emotionally flat, reveals surprisingly elevated activeness during non-overlap—especially in the mouth and upper face—likely reflecting articulatory dynamics. During overlapping speech, neutral conditions still show modest yet notable expressivity in the brow and mouth, exceeding those seen in happy or angry overlap, perhaps reflecting compensatory modulation to maintain clarity.

Together, these results highlight how both emotion and interactional structure shape the spatial deployment of expressive motion: non-overlapping speech fosters distributed engagement across face and hands, while overlapping speech selectively concentrates motion in high-arousal facial zones (e.g., brow) and attenuates peripheral expressivity—especially in gesture.

## 4.2 Facial and hand activeness by speech conditions and emotions

To statistically evaluate the differences in the average framewise displacement magnitudes under each emotion and speech condition differences, we conducted repeated-measures ANOVA on region-level activeness values (Figure 6). Across nearly all regions, non-overlapping speech elicited significantly greater activeness compared to overlapping speech, most prominently in the mouth (Figure 6c) ($F(1,132) = 229.49$, $p < .001$, $\eta^2 = .258$), lower face (Figure 6f) ($F = 217.92$, $p < .001$, $\eta^2 = .249$), and middle face (Figure 6e) ($F = 175.29$, $p < .001$, $\eta^2 = .203$). This consistent enhancement suggests that turn-taking provides expressive bandwidth, enabling speakers to coordinate emotion-rich facial and bodily signals more effectively. Rather than serving as mere reactionary outputs, these expressions appear co-constructed with interactional timing, highlighting the interdependence of structure and emotion.

Emotion category also significantly influenced regional motion, but in a more nuanced and spatially distinct fashion. While happy and neutral expressions were generally more expressive than angry, it was sadness that yielded especially notable activation in non-overlapping speech—particularly in the lower face ($F = 76.75$, $p < .001$, $\eta^2 = .133$) (Figure 6f) and mouth ($F = 71.27$, $p < .001$, $\eta^2 = .126$) (Figure 6c). This counters prevailing assumptions that sadness is behaviorally "muted," revealing that given space and timing, low-arousal emotions can manifest in rich, sustained

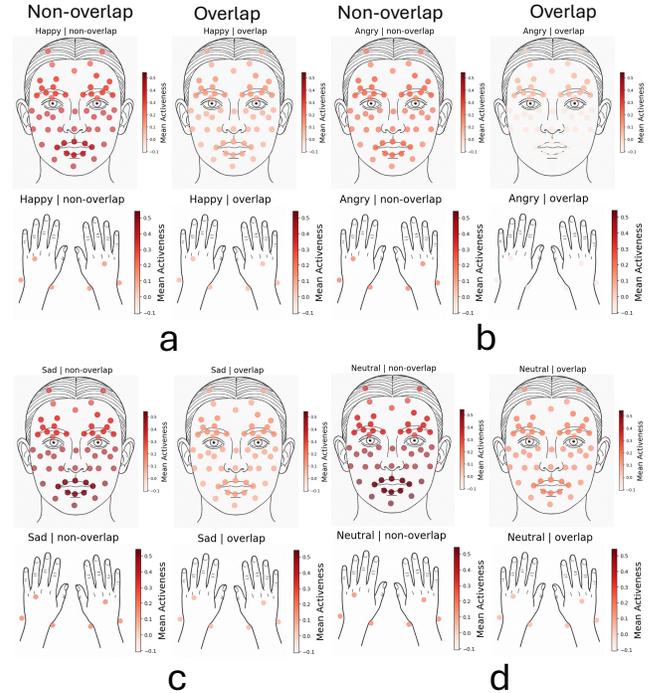

**Figure 5: Marker-level heatmaps of mean framewise activeness across facial and hand regions for each emotion (Happy (a), Angry (b), Sad (c), and Neutral (d)) under non-overlapping (left panels) and overlapping (right panels) speech conditions. Warmer colors indicate higher displacement magnitudes.**

motion, especially around articulators and midfacial muscles involved in subtle affect display.

In contrast, anger during overlapping speech was marked by a compression of expression: both the face and hands showed reduced activeness, with residual motion magnitude localized in the brow and forehead areas. This pattern likely reflects tension or internal regulation in high-conflict, high-arousal moments, where overlapping speech disrupts the full-body synchrony typically associated with expressive states. Even neutral states exhibited non-trivial activeness in middle and lower facial regions—particularly during non-overlapping segments. These findings suggest that "neutral" is not void of expression, but reflects articulatory and interactional dynamics intrinsic to natural speech. In effect, so-called neutral expressions are shaped as much by the rhythm of conversation as by the absence of overt emotion.

Motion in the eyebrows ($F = 10.24$, $p < .001$, $\eta^2 = .022$) (Figure smaller, emerged in regions such as the mouth ($F = 2.77$, $p = .041$) (Figure 6c), lower face ($F = 2.97$, $p = .032$) (Figure 6f), and hands ($F = 4.42$, $p = .004$) (Figure 6h). These effects underscore that the expressive role of each region is not fixed, but contextually modulated by both affective intent and dialogic flow. For instance, sadness may involve more gestural involvement when the speaker has uninterrupted space, while angry expressions may be tightly held in overlapping, competitive speech. Together, these results underscore a central insight: emotional expression is not merely a function of internal state, but of how interactional affordances allow that state to unfold across the body. Effective emotion modeling—whether in human perception or artificial systems—must consider not only what is being expressed, but how conversational timing and structure shape its emergence.

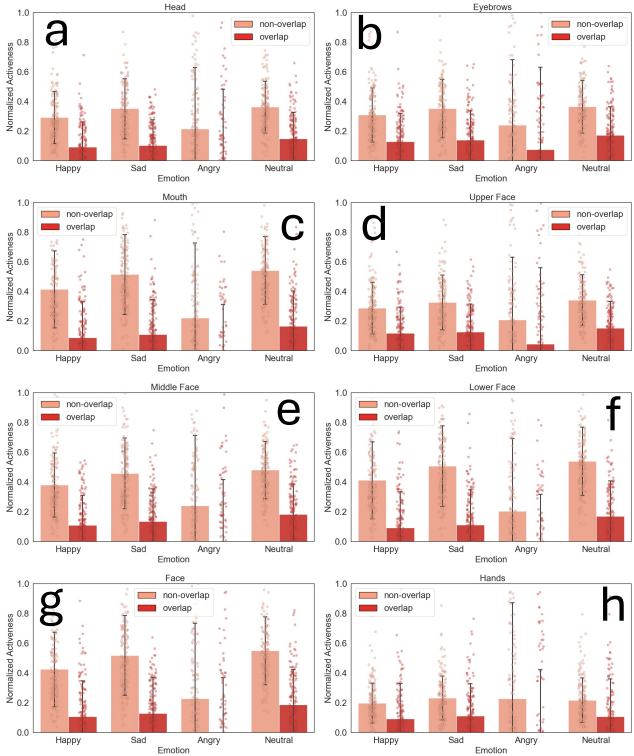

Figure 6: Mean activeness for each region—head (a), eyebrows (b), mouth (c), upper face (d), middle face (e), lower face (f), total face (g), and hands (h)—segmented by emotion category and speech condition.

## 4.3 Speech-to-motion mapping performance across facial and gestural regions

We evaluated how well different acoustic feature sets predicted framewise facial and hand motion using the Affine Minimum Mean Square Error (AMMSE) estimator, summarized in Figure 7. Across all motion regions, mapping accuracy—quantified by mean Pearson correlation (r)—was consistently highest when using prosodic features (e.g., pitch, energy), followed closely by and spectral features (MFCCs) and outperformed emotion-derived descriptors (arousal, valence), both in absolute accuracy and regional consistency. For total facial motion, prosody yielded the highest predictive accuracy (r = .47, SEM = .006), with MFCCs closely trailing (r = .44, SEM = .006). In contrast, arousal- and valence-based features showed weaker predictive power, with correlations peaking at r = 0.33 for arousal (mouth) and r = 0.31 for valence (lower face), and falling below r = 0.25 in eyebrow, hand, and head regions. This discrepancy likely reflects the functional granularity of the features. While prosody and MFCCs preserve moment-to-moment spectral and rhythmic structure—closely tied to articulatory and gestural timing—arousal and valence estimates are higher-level abstractions, often smoothed over time. Their reduced sensitivity to short-term dynamics may explain their diminished alignment with real-time motion.

The most predictable regions were the lower face and mouth, where prosody-based mappings reached r = .46, and MFCC-based mappings followed at r = .44 and .43, respectively. These regions are deeply involved in speech articulation and likely reflect tight biomechanical coupling with vocal production. Such strong audio-to-motion mapping supports the hypothesis that articulatory gestures are co-regulated by both phonetic and affective speech.

Motion in the middle face, including cheeks and nasal markers, was also well-predicted by prosody (r = .41) and MFCCs (r = .39). This finding is particularly noteworthy, as the midface has been underexplored in audiovisual modeling despite its role in conveying nuanced emotional and conversational signals. The moderate accuracy here suggests that midface dynamics are partially governed by acoustic structure, possibly reflecting subtle co-articulatory or emotive modulation.

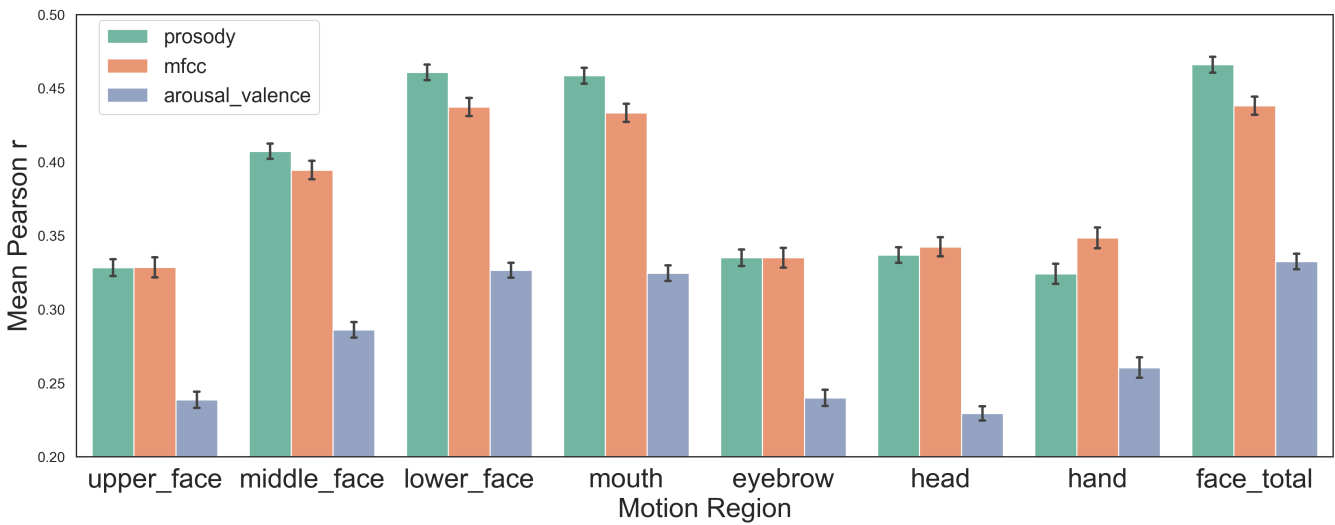

Figure 7: Speech-to-motion mapping AMMSE prediction accuracy.

In contrast, the upper face and eyebrows—regions traditionally linked to high-level cognitive and social cues—showed lower mapping performance, with prosody and MFCCs plateauing at r = .33–.34, and arousal/valence features dropping further to r = .24. This discrepancy underscores a functional divide: whereas lower and midface regions are tightly coupled with speech acoustics, upper facial movements may emerge more independently, modulated by discourse intent, mental state, or social signaling demands.

The hand and head regions, though distal from the vocal tract, still displayed nontrivial correlations (e.g., MFCC-hand: r = 0.35; prosody-head: r = 0.34), pointing to temporal entrainment between vocal and motor systems. These findings suggest that gestural timing is not arbitrary, but aligned with acoustic salience, reinforcing theories of prosody–gesture synchronization. Still, performance was notably lower when using arousal/valence features (r = .23 for head, r = .26 for hand), reaffirming that such affective embeddings, while informative for semantic interpretation, are less effective for frame-level motor prediction.

Taken together, these results reveal a topographic and feature-based dissociation in speech-to-motion coupling. Acoustic features that encode low-level dynamics (prosody, MFCCs) are better suited for predicting expressive motion than abstract emotion dimensions, particularly in regions involved in articulation and rhythmic coordination. The relatively weak performance of arousal and valence models—despite their psychological interpretability—raises critical implications for affective computing: emotion recognition pipelines may need to integrate fine-grained signal-level cues to capture the temporal precision required for real-time multimodal modeling.

Figure 8 illustrates how audiovisual coupling varies across affective states and acoustic features using AMMSE-based correlation maps. In high arousal (a) and high valence (c) conditions, facial–speech alignment is strongest in the middle and lower face—particularly the mouth—with minimal difference between speech conditions, indicating stable coupling under positive or intense affect. Conversely, low arousal (b) and low valence (d) show stronger face and hand correlations during overlapping speech, especially for low arousal, where a clear gradient emerges from weak upper-face to strong lower-face and hand synchrony. This suggests that under subdued affect, speakers compensate for vocal interference with broader, distributed nonverbal expressivity.

Interestingly, only arousal—but not valence—enhances hand coupling, pointing to arousal's greater role in modulating motor entrainment. MFCC (e) and prosodic (f) features outperform emotion-based descriptors across all regions, especially under overlap, where r-values exceed 0.4 in the mouth, cheeks, and hands. These results reveal a functional hierarchy: low-level acoustic

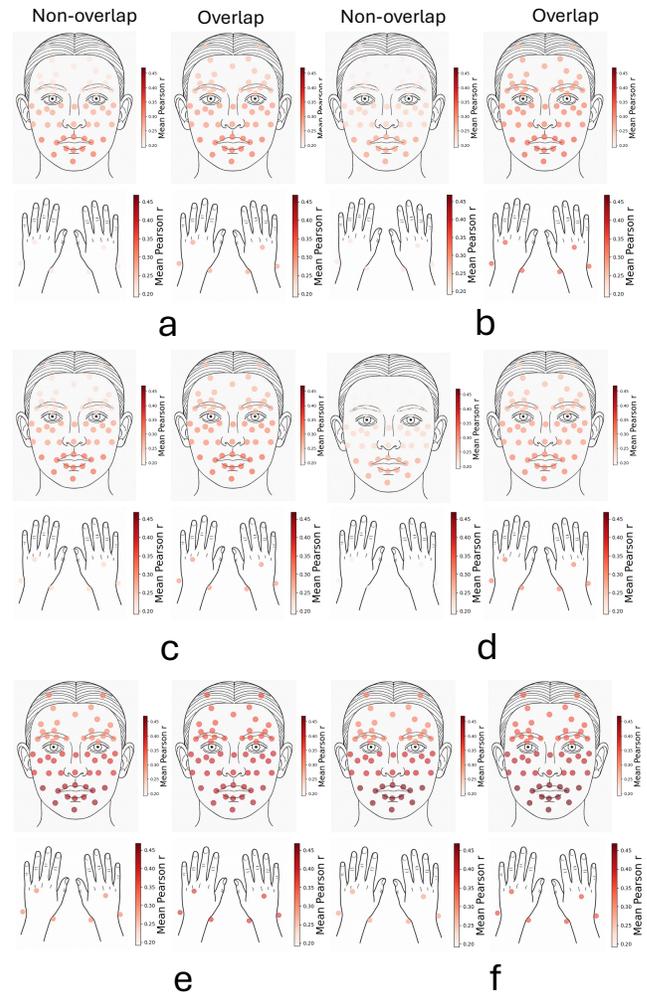

Figure 8: Anatomical heatmaps of AMMSE correlation strength across acoustic feature sets, affective dimensions, and speech conditions. High arousal (a), Low arousal (b), High valence (c), Low valence (d), MFCC (e), prosody (f)

features, rich in temporal dynamics, drive robust speech–motion alignment, while abstract emotion features are less predictive. Crucially, arousal amplifies this coupling more than valence, especially when conversational structure introduces expressive challenges.

## 5 Conclusion

This study advances the understanding of multimodal emotional expression by demonstrating how vocal emotion signals dynamically align with facial and hand movements in dyadic interaction. By incorporating both speech-driven features (prosody, MFCCs) and model-predicted emotional dimensions (arousal, valence), we show that expressive behavior is jointly shaped by affective state, acoustic dynamics, and conversational structure. Our analysis revealed that non-overlapping speech consistently enhances facial and gestural activeness, supporting smoother, more coordinated multimodal expression. Conversely, overlapping speech dampens peripheral gestures while concentrating motion in high-arousal facial zones, particularly the brow—highlighting a strategic redistribution of expressive effort under competitive interaction. From a mapping perspective, prosody and MFCCs outperformed emotion-based features in predicting real-time motion, especially in regions tied to articulation (mouth, lower face). Emotion-derived descriptors—while interpretable—showed weaker coupling, with arousal playing a more central role than valence in driving synchrony, especially under challenging (overlap) conditions. These findings point to a functional hierarchy of acoustic-to-motion coupling, where low-level temporal features better capture embodied synchrony. They also underscore the need for emotion recognition models to move beyond static or isolated modalities, embracing temporally aligned, multimodal signals that reflect the fluid and embodied nature of real-world affective interaction.

## Acknowledgments

This work was supported by the JST Moonshot R&D, Japan (Grant Number: JPMJMS2292).

## References

[1] David McNeill. 1992. *Hand and Mind: What Gestures Reveal about Thought*. University of Chicago Press, Chicago.
[2] Adam Kendon. 2004. Gesture: Visible Action as Utterance. Cambridge University Press.
[3] Janet Beavin Bavelas, Nicole Chovil, Linda Coates, and Lori Roe. 1995. Gestures Specialized for Dialogue. Personality and Social Psychology Bulletin 21, 4 (1995), 394–405. DOI: https://doi.org/10.1177/0146167295214010.
[4] Soujanya Poria, Erik Cambria, Rajiv Bajpai, and Amir Hussain. 2017. A review of affective computing: From unimodal analysis to multimodal fusion. Information Fusion 37 (2017), 98–125. DOI: https://doi.org/10.1016/j.inffus.2017.02.003
[5] Zhihong Zeng, Maja Pantic, Glenn I. Roisman, and Thomas S. Huang. 2009. A survey of affect recognition methods: Audio, visual, and spontaneous expressions. *IEEE Transactions on PAMI* 31, 1 (2009), 39–58. DOI: https://doi.org/10.1109/TPAMI.2008.52
[6] Carlos Busso et al. 2004. Analysis of emotion recognition using facial expressions, speech and multimodal information. *Proceedings of ICMI* (2004), 205–211. DOI: https://doi.org/10.1145/1027933.1027968.
[7] Klaus R. Scherer and Harald Ellgring. 2007. Multimodal expression of emotion: Affect programs or componential appraisal patterns? *Emotion* 7, 1 (2007), 158–171. DOI: https://doi.org/10.1037/1528-3542.7.1.158.
[8] Tadas Baltrušaitis, Chaitanya Ahuja, and Louis-Philippe Morency. 2019. Multimodal machine learning: A survey and taxonomy. IEEE TPAMI 41, 2 (2019), 423–443. DOI: https://doi.org/10.1109/TPAMI.2018.2798607.
[9] Michael Kipp and Jean-Claude Martin. 2009. Gesture and emotion: Can basic gestural form features discriminate emotions? *ACII* (2009), 1–8. DOI: https://ieeexplore.ieee.org/document/5349544
[10] Susan Goldin-Meadow. 2003. *Hearing Gesture: How Our Hands Help Us Think*. Harvard University Press.
[11] Paolo Bernardis and Maurizio Gentilucci. 2006. Speech and gesture share the same communication system. Neuropsychologia 44, 2 (2006), 178–190. DOI: https://doi.org/10.1016/j.neuropsychologia.2005.05.007.
[12] Cravotta, A., Busà, M.G., & Prieto, P. (2019). Effects of Encouraging the Use of Gestures on Speech. Journal of Speech, Language, and Hearing Research, 62(9), 3204-3219. DOI: https://doi.org/10.1044/2019_JSLHR-S-18-0493.
[13] Stewart, J.R., Vigil, D.C., Olszewski, A., & Thornock, C. (2022). Gesture use in children with autism spectrum disorder: a scoping review. Clinical Archives of Communication Disorders, 7(3), 94–104. DOI: https://doi.org/10.21849/cacd.2022.00675.
[14] Jenkins, T., & Pouw, W. (2023). Gesture–speech coupling in persons with aphasia: A kinematic-acoustic analysis. Journal of Experimental Psychology: General, 152(5), 1469–1483. DOI: https://doi.org/10.1037/xge0001346.
[15] Kita, S., van Gijn, I., & van der Hulst, H. (1998). Movement phases in signs and co-speech gestures, and their transcription by human coders. In I. Wachsmuth & M. Fröhlich (Eds.), *Gesture and sign language in human-computer interaction: International gesture workshop Bielefeld, Germany, September 17–19, 1997: Proceedings* (Lecture Notes in Computer Science, Vol. 1371, pp. 23-35). Springer. ISBN 9783540644248
[16] Löffler, D., Schmidt, N., & Tscharn, R. (2018). Multimodal Expression of Artificial Emotion in Social Robots Using Color, Motion and Sound. In *Proceedings of the 2018 ACM/IEEE International Conference on Human-Robot Interaction* (pp. 334–343). ACM/IEEE. DOI: https://doi.org/10.1145/3171221.3171261.
[17] Morency, L.-P., Sidner, C., Lee, C., & Darrell, T. (2005). Contextual recognition of head gestures. In *Proceedings of the 7th International Conference on Multimodal Interfaces* (pp. 18–24). ACM. DOI: https://doi.org/10.1145/1088463.1088470.
[18] Agustín Gravano and Julia Hirschberg. 2011. Turn-taking cues in task-oriented dialogue. *Computer Speech & Language* 25, 3 (2011), 601–634. DOI: https://doi.org/10.1016/j.csl.2010.10.003.
[19] Mattias Heldner and Jens Edlund. 2010. Pauses, gaps and overlaps in conversations. *Journal of Phonetics* 38, 4 (2010), 555–568. DOI: https://doi.org/10.1016/j.wocn.2010.08.002.
[20] Alex Pentland. 2008. *Honest Signals: How They Shape Our World*. MIT Press. https://doi.org/10.7551/mitpress/8022.001.0001
[21] Khiet P. Truong and David A. van Leeuwen. 2007. Automatic discrimination between laughter and speech. *Speech Communication* 49, 2 (2007), 144–158. DOI: https://doi.org/10.1016/j.specom.2007.01.001.
[22] Carlos Busso and Shrikanth Narayanan. 2007. Interrelation between speech and facial gestures in emotional utterances: A single subject study. *IEEE Transactions on Audio, Speech, and Language Processing* 15, 8 (2007), 2331–2347. DOI: https://doi.org/10.1109/TASL.2007.902008.
[23] Gunes, H., & Pantic, M. (2010). Automatic, Dimensional and Continuous Emotion Recognition. *International Journal of Synthetic Emotions, 1*(1), 68–99. DOI: https://doi.org/10.4018/jse.2010101605.
[24] Sadoughi, N., & Busso, C. (2019). Speech-driven animation with meaningful behaviors. *Speech Communication, 110*, 90–100. DOI: https://doi.org/10.1016/j.specom.2019.04.005.
[25] Busso, C., et al. (2008). IEMOCAP: Interactive emotional dyadic motion capture database. *Language Resources and Evaluation, 42*(4), 335–359. DOI: https://doi.org/10.1007/s10579-008-9076-6.
[26] Yang et al. (2021). SUPERB: Speech processing Universal PERformance Benchmark. In *22nd Annual Conference of the International Speech Communication Association, INTERSPEECH 2021* (pp. 3161-3165). (Proceedings of the Annual Conference of the International Speech Communication Association, INTERSPEECH; Vol. 4). International Speech Communication Association. https://doi.org/10.21437/Interspeech.2021-1775
[27] Baevski, A., Zhou, H., Mohamed, A., & Auli, M. (2020). wav2vec 2.0: A framework for self-supervised learning of speech representations. arXiv:2006.11477 [cs.SD]. https://arxiv.org/abs/2006.11477
[28] Wagner, J., Triantafyllopoulos, A., Wierstorf, H., Schmitt, M., Burkhardt, F., Eyben, F., & Schuller, B. W. (2022). Dawn of the transformer era in speech emotion recognition: closing the valence gap. *arXiv:2203.07378 [eess.AS]*. https://arxiv.org/abs/2203.07378